\documentclass[journal]{IEEEtran}  

\IEEEoverridecommandlockouts                              

\usepackage[mode=tex]{standalone} 

\usepackage{cite}
\usepackage{color}

\usepackage{verbatim}
\usepackage{booktabs}
\usepackage{amsmath}

\usepackage{graphicx}
\usepackage{tabu}
\usepackage[font={footnotesize}]{caption}
\usepackage{array, ltablex, multirow,ragged2e}
\usepackage{hyperref}
\usepackage{longtable}
\usepackage{mdframed}
\usepackage{subfigure}
\usepackage{dsfont}
\usepackage{float}
\usepackage{algorithm}
\usepackage[noend]{algpseudocode}
\usepackage{tikz,pgfplots,tkz-graph}
\usepgfplotslibrary{fillbetween}

\makeatletter
\def\BState{\State\hskip-\ALG@thistlm}
\makeatother

\usepackage{amsmath} 
\usepackage{amssymb}  
\usepackage{amsthm}
\usepackage{bm}
\theoremstyle{plain}
\newtheorem{thm}{Theorem}
\theoremstyle{remark}
\newtheorem{rem}[thm]{Remark}
\theoremstyle{plain}
\newtheorem{prop}{Proposition}
\graphicspath{{../figures/}}

\pgfplotsset{compat=newest}%
\usetikzlibrary{positioning,matrix,shapes.multipart,shapes.misc,spy}%
\tikzstyle{annotation}=[fill=white]%

\newcommand{\vect}[1]{\mathbf{#1}}
\newcommand{\ptx}{\tau}
\newcommand{\tx}{f_{\ptx}}
\newcommand{\prx}{\rho}
\newcommand{\rx}{\mathbf{f}_{\prx}}
\newcommand{\transpose}{\intercal}

\newcommand{\iter}{N}
\newcommand{\itertx}{N_T}
\newcommand{\iterrx}{N_R}

\usetikzlibrary{shapes.arrows}

\tikzset{%
	partial ellipse/.style args={#1:#2:#3}{%
		insert path={+ (#1:#3) arc (#1:#2:#3)}%
	}%
}%
\tikzstyle{style A}=[black, mark=diamond*, mark options={solid, fill=white, mark size=2.0pt}, solid]%
\tikzstyle{style B}=[color=blue, mark=*, mark options={solid, fill=white, mark size=1.5pt}, dotted]%
\tikzstyle{annotation}=[fill=white]%

\tikzstyle{diamond marker}=[mark=diamond*, mark options={solid, fill=white, mark size=2.0pt}]
\tikzstyle{triangle marker}=[mark=triangle*, mark options={solid, fill=white, mark size=2.0pt}]
\tikzstyle{square marker}=[mark=square*, mark options={solid, fill=white, mark size=1.3pt}]
\tikzstyle{circle marker}=[mark=*, mark options={solid, fill=white, mark size=1.5pt}]

\newcommand{\RevA}[1]{%
	{\color{black}%
	#1%
	}%
}%

\newcommand{\RevB}[1]{%
	{\color{black}%
	#1%
	}%
}%

\definecolor{DarkGreen}{rgb}{0.0, 0.6, 0.0}
\newcommand{\RevC}[1]{%
    {\color{black}
	#1%
	}%
}%

\DeclareMathOperator*{\argmax}{arg\,max}

\title{Learning Physical-Layer Communication\\ with Quantized Feedback}

\author{%
Jinxiang Song, 
Bile Peng, 
Christian H\"{a}ger, 
Henk Wymeersch, Anant Sahai
\thanks{
J.~Song, B.~Peng, C.~H\"ager, and H.~Wymeersch are with the Department of Electrical Engineering, Chalmers University of Technology, Gothenburg, Sweden. email: jinxiang@student.chalmers.se, \{bile.peng, christian.haeger, henkw\}@chalmers.se. C.~H\"ager is also with the Department of Electrical and Computer 
Engineering, Duke University, Durham, USA. A.~Sahai is with the Department of Electrical Engineering and Computer Science, UC Berkeley, Berkeley, USA. 
The work of C.~H\"ager was supported by the European Union's Horizon 2020 research and innovation programme under the Marie Sk\l{}odowska-Curie grant No.~749798. H.~Wymeersch was supported by the Swedish Research Council under grant No.~2018-03701. 
}
}

\begin{document}

\maketitle

\begin{abstract}
Data-driven optimization of transmitters and receivers can reveal new modulation and detection schemes and enable physical-layer communication over unknown channels. Previous work has shown that practical implementations of this approach require a feedback signal from the receiver to the transmitter. In this paper, we study the impact of quantized feedback on data-driven learning of physical-layer communication. A novel quantization method is proposed, which exploits the specific properties of the feedback signal and is suitable for non-stationary signal distributions. The method is evaluated for linear and nonlinear channels. Simulation results show that feedback quantization does not appreciably affect the learning process and can lead to \RevA{similar performance as compared to the case where unquantized feedback is used for training}, even with $1$-bit quantization. In addition, it is shown that learning is surprisingly robust to noisy feedback where random bit flips are applied to the quantization bits. 
\end{abstract}

\section{Introduction}

As communication systems become more complex, \RevA{physical-layer design, i.e., devising optimal transmission and detection methods,} has become harder as well. This is true not only in wireless communication, where hardware impairments and quantization have increasingly become a limitation on the achievable performance, but also in optical communication, for which the nonlinear nature of the channel precludes the use of standard approaches. This has led to a new line of research \RevA{on physical-layer communication} where transmission and detection methods are learned from data. The general idea is to regard the transmitter and receiver as parameterized functions (e.g., neural networks) and find good parameter configurations using large-scale gradient-based optimization approaches from machine learning. 

Data-driven methods have mainly focused on learning receivers assuming a given transmitter and channel, e.g., for MIMO detection \cite{Samuel2017} or decoding \cite{Nachmani2018}. These methods have led to algorithms that either perform better or exhibit lower complexity than model-based algorithms. More recently, end-to-end learning of both the transmitter and receiver has been proposed for various \RevA{physical-layer} applications including wireless \cite{OShea2017, Doerner2018}, nonlinear optical \cite{karanov2018end, li2018achievable, Jones2018}, and visible light communication\cite{Lee2018}.

In practice, gradient-based transmitter optimization is problematic since it requires a known and differentiable channel model. One approach to circumvent this limitation is to first learn a surrogate channel model, e.g., through an adversarial process, and use the surrogate model for the optimization \cite{OShea2018, Ye2018}. We follow a different approach based on stochastic transmitters, where the transmitted symbol for a fixed message is assumed to be a random variable during the training process \cite{Aoudia2018, Aoudia2018a, DeVrieze2018}. This allows for the computation of \emph{surrogate gradients} which can then be used to update the transmitter parameters. A related approach is proposed in \cite{Raj2018}.\footnote{See \cite[Sec.~III-C]{Aoudia2018a} for a discussion about the relationship between the approaches in \cite{Aoudia2018, Aoudia2018a, DeVrieze2018} and \cite{Raj2018}.}

In order to compute the surrogate gradients, the transmitter must receive a \emph{feedback signal} from the receiver. This feedback signal can either be perfect \cite{Aoudia2018, Aoudia2018a, Raj2018, DeVrieze2018} or noisy \cite{Goutay2018}. \RevC{In the latter case, it was proposed in \cite{Goutay2018} to regard the feedback transmission as a separate communication problem for which optimized transmitter and receiver pairs can again be learned. The proposed training scheme in \cite{Goutay2018} alternates between optimizing the different transmitter/receiver pairs, with the intuition that training improvements for one pair lead to better training of the other pair (and vice versa). Thus, both communication systems improve simultaneously and continuously until some predefined stopping criterion is met (see Alg.~3 in \cite{Goutay2018}).  The assumed feedback link in \cite{Goutay2018} only allowed for the transmission of real numbers over an additive white Gaussian noise (AWGN) channel.} In practice, however, signals will be quantized to a finite number of bits, including the feedback signal. To the best of our knowledge, such quantization has not yet been considered in the literature. Studies on quantization have been conducted so far only in terms of the transmitter and receiver processing, for example when the corresponding learned models are implemented with finite resolution \cite{Kim2018b, Tang2018, Teng2018, Fougstedt2018ecoc, Aoudia2019}. 

In this paper, we analyze the impact of quantization of the feedback signal on data-driven learning of physical-layer communication over an unknown channel. \RevC{Compared to \cite{Goutay2018}, the feedback transmission scheme is not learned. Instead, we show that due to the specific properties of the feedback signal, an adaptive scheme based on simple pre-processing steps followed by a fixed quantization strategy can lead to} \RevA{similar performance as compared to the case where unquantized feedback is used for  training, even with $1$-bit quantization.} We provide a theoretical justification for the proposed approach and perform extensive simulations for both linear Gaussian and \RevC{nonlinear phase-noise channels}. \RevB{The detailed contributions in this paper are as follows: 
\begin{enumerate}
    \item We propose a novel quantization method for feedback signals in data-driven learning of physical-layer communication. The proposed method addresses a major shortcoming in previous work, in particular the assumption in \cite{Goutay2018} that feedback losses can be transmitted as unquantized real numbers over an AWGN channel. 
    
    \item We conduct a thorough numerical study demonstrating the effectiveness of the proposed scheme. We investigate the impact of the number of quantization bits on the performance and the training process, showing that $1$-bit quantization can provide performance similar to unquantized feedback. In addition, it is shown that the scheme is robust to noisy feedback where the quantized signal is perturbed by random bit flips. 
    
    \item We provide a theoretical justification for the effectiveness of the proposed approach in the form of Propositions 1 and 2. In particular, it is proved that feedback quantization and bit flips manifest themselves merely as a scaling of the expected gradient used for parameter training. Moreover, upper bounds on the variance of the gradient are derived in terms of the Fisher information matrix of the transmitter parameters. 
    
\end{enumerate}
}%

\subsubsection*{Notation}
Vectors will be denoted with lower case letters in bold (e.g., $\mathbf{x}$), with $x_n$ or $[\mathbf{x}]_n$ referring to the $n$-th entry in $\mathbf{x}$; matrices will be denoted in bold capitals (e.g., $\mathbf{X}$); $\mathbb{E}(\{\mathbf{x}\}$ denotes the expectation operator; $\mathbb{V}(\mathbf{x})$ denotes  the variance (the trace of the covariance matrix) of the random vector $\mathbf{x}$ (i.e., $\mathbb{V}\{\mathbf{x}\}=\mathbb{E}\{\mathbf{x}^\transpose\mathbf{x}\}-(\mathbb{E}\{\mathbf{x}\})^\transpose(\mathbb{E}\{\mathbf{x}\})$).

\begin{figure}
\centering
\includegraphics[width=0.95\columnwidth]{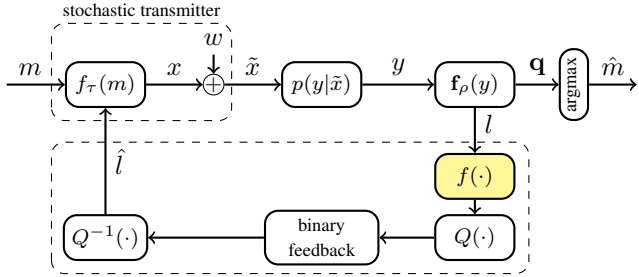}
\caption{Data-driven learning model where the discrete time index $k$ (e.g., $m_k$) is omitted for all variables. The quantization and binary feedback is shown in the lower dashed box, while the proposed pre-processor is highlighted. Note that $w=0$ for the receiver learning (Sec.~\ref{sec:receiver_learning}).}
\label{fig:model}
\end{figure}
\section{System Model}
\label{sec:model}

\newcommand{\define}{\triangleq}

We wish to transmit messages \RevC{$m \in \{1, \ldots, M\}$} over an a priori unknown \RevB{static} memoryless channel which is defined by a conditional probability density function (PDF) $p(y|x)$, where $x,y \in \mathbb{C}$ and $M$ is the total number of messages.\footnote{In this paper, we restrict ourselves to two-dimensional (i.e., complex-valued) channel models, where the generalization to an arbitrary number of dimensions is straightforward.} 
The communication system is implemented by representing the transmitter and receiver as two parameterized functions $\tx : \RevC{\{1, \ldots, M\}} \to \mathbb{C}$ and $\rx : \mathbb{C} \to [0,1]^M$,  where \RevC{$[a,b]^M$ is the $M$--fold Cartesian product of the $[a,b]$--interval (i.e., the elements in $[a,b]^M$ are vectors of length $M$ with entries between $a$ and $b$ inclusively) and } $\ptx$ and $\prx$ are sets of transmitter and receiver parameters, respectively. The transmitter maps the $k$-th message $m_k$ to a complex symbol $x_k = \tx(m_k)$, where an average power constraint according to $\mathbb{E}\{ |x_k|^2\} \le P$ is assumed. The symbol $x_k$ is sent over the channel and the receiver maps the channel observation $y_k$ to a probability vector $\vect{q}_k = \rx(y_k)$, where one may interpret the components of $\vect{q}_k$ as estimated posterior probabilities for each possible message. Finally, the receiver outputs an estimated message according to $\hat{m}_k = \arg \max_m [\vect{q}_k]_m$, where $[\vect{x}]_{m}$ returns the $m$-th component of $\vect{x}$. The setup is depicted in the top branch of the block diagram in Fig.~\ref{fig:model}, where the random perturbation $w$ in the transmitter can be ignored for now.

We further assume that there exists a feedback link from the receiver to the transmitter, which, as we will see below, facilitates transmitter learning. In general, our goal is to learn optimal transmitter and receiver mappings $\tx$ and $\rx$ using limited feedback. 

\section{Data-Driven Learning}
\label{sec_theory}

\newcommand{\ExpL}{\ell} 
\newcommand{\EmpL}{\ell^{\text{e}}} 

In order to find good parameter configurations for $\ptx$ and $\prx$, a suitable optimization criterion is required. Due to the reliance on gradient-based methods, conventional criteria such as the symbol error probability $\Pr(m_k \neq \hat{m}_k)$ cannot be used directly. Instead, it is common to minimize the expected cross-entropy loss defined by
\begin{align}
    \ExpL (\tau,\rho) \define - \mathbb{E}\{\log([\rx(y_k)]_{m_k})\},
\end{align}
where the dependence of $\ExpL (\tau,\rho)$ on $\tau$ is implicit through the distribution of $y_k$. 

A major practical hurdle is the fact that the gradient $\nabla_\tau \ExpL (\tau,\rho)$ cannot actually be evaluated because it requires a known and differentiable channel model. To solve this problem, we apply the alternating optimization approach proposed in \cite{Aoudia2018, Aoudia2018a}, which we briefly review in the following. For this approach, one alternates between optimizing first the receiver parameters $\prx$ and then the transmitter parameters $\ptx$ for a certain number of iterations $\iter$. To that end, it is assumed that the transmitter and receiver share common knowledge about a database of training data $m_k$.

\subsection{Receiver Learning}
\label{sec:receiver_learning}

For the receiver optimization, the transmitter parameters $\ptx$ are assumed to be fixed. The transmitter maps a mini-batch of uniformly random training messages $m_k$, \RevC{$k \in \{1,\ldots, B_R\}$}, to symbols satisfying the power constraint and transmits them over the channel. The receiver observes $y_1,\ldots,y_{B_R}$ and generates $B_R$ probability vectors $\rx(y_1), \ldots, \rx(y_{B_R})$. 

The receiver then updates its parameters $\prx$ according to $\prx_{i+1} = \prx_{i} - \alpha_R \nabla_{\prx} \EmpL_R(\prx_i)$, where
\begin{align}
     \EmpL_R(\prx) = -\frac{1}{B_R}\sum^{B_R}_{k=1}\log([\rx(y_{k})]_{m_{k}})
\end{align}
is the empirical cross-entropy loss associated with the mini-batch and $\alpha_R$ is the learning rate. This procedure is repeated iteratively for a fixed number of iterations $\iterrx$.

\subsection{Transmitter Learning}
For the transmitter optimization, the receiver parameters are assumed to be fixed. The transmitter generates a mini-batch of uniformly random training messages $m_k$, \RevC{$k \in \{1,\ldots, B_T\}$}, and performs the symbol mapping as before. However, before transmitting the symbols over the channel, a small Gaussian perturbation is applied, which yields $\tilde{x}_k = x_k + w_k$, where $w_k \sim \mathcal{CN}(0,\sigma_p^2)$ and reasonable choices for $\sigma_p^2$ are discussed in Sec.~\ref{sec:simulation}. Hence, we can interpret the transmitter as stochastic, described by the PDF
\begin{align}
    \label{eq:gaussian_policy}
    \pi_{\ptx}(\tilde{x}_k|m_k) = \frac{1}{\pi \sigma_p^2}
    \exp \left(
        - \frac{|\tilde{x}_k - \tx(m_k) |^2}{\sigma_p^2}
    \right).
\end{align}
Based on the received channel observations, the receiver then computes per-sample losses $l_k=-\log([\rx(y_{k})]_{m_{k}}) \in \mathbb{R}$ for \RevC{$k \in \{1,\ldots,B_T\}$}, and feeds these back to the transmitter via the feedback link. The corresponding received losses are denoted by 
$\hat{l}_k$, where ideal feedback corresponds to $\hat{l}_k = l_k$. Finally, the transmitter updates its parameters $\ptx$ according to $\ptx_{i+1} = \ptx_{i} - \alpha \nabla_{\ptx} \EmpL_T(\ptx_i)$, where
\begin{align}
    \nabla_{\ptx} \EmpL_T(\ptx) = \frac{1}{B_T}\sum_{k=1}^{B_T} \hat{l}_k \nabla_{\ptx} \log \pi_{\ptx}(\tilde{x}_k|m_k).  \label{eq:PolicyGradient1}         
\end{align}
This procedure is repeated iteratively for a fixed number of iterations $\itertx$, after which the alternating optimization continues again with the receiver learning. The total number of gradient steps in the entire optimization is given by $\iter(\itertx+\iterrx)$.

A theoretical justification for the gradient in \eqref{eq:PolicyGradient1} can be found in  \cite{Aoudia2018, Aoudia2018a, DeVrieze2018}. In particular, it can be shown that the gradient of $\ExpL_T(\ptx) = \mathbb{E}\left\{  l_k \right\}$ is given by
\begin{align}
    \label{eq:policy_gradient}
    \nabla_{\ptx} \ExpL_T(\ptx) = \mathbb{E}\left\{ l_k \nabla_{\ptx} \log \pi_{\ptx}(\tilde{x}_k|m_k)\right\}, 
\end{align}
where the expectations are over the message, transmitter, and channel distributions. Note that \eqref{eq:PolicyGradient1} is the corresponding sample average for finite mini-batches assuming $\hat{l}_k = l_k$. 

\begin{rem}
    As pointed out in previous work, the transmitter optimization can be regarded as a simple form of reinforcement learning. In particular, one may interpret the transmitter as an agent exploring its environment according to a stochastic exploration policy defined by \eqref{eq:gaussian_policy} and receiving (negative) rewards in the form of per-sample losses. The state is the message $m_k$ and the transmitted symbol $\tilde{x}_k$ is the corresponding action. The learning setup belongs to the class of \emph{policy gradient methods}, which rely on optimizing parameterized policies using gradient descent. We will make use of the following well-known property of policy gradient learning:\footnote{To see this, one may first apply 
$\nabla_{\ptx} \log \pi_{\ptx} = \frac{\nabla_{\ptx} \pi_{\ptx} }{\pi_{\ptx}}$ and then use the fact that $\int \nabla_{\ptx}\pi_{\ptx}(\tilde{x}|m) \text{d} \tilde{x} = 0$ since $\int \pi_{\ptx}(\tilde{x}|m) \text{d} \tilde{x} = 1$.}
\begin{align}
    \label{eq:exp_grad_log_policy}
    \mathbb{E}\left\{ \nabla_{\ptx} \log \pi_{\ptx}(\tilde{x}_k|m_k)\right\} = 0.
\end{align}
\end{rem}

\subsection{Loss Transformation}
\label{sec:losstransform}

\RevA{The per-sample losses can be transformed through a pre-processing function $f: \mathbb{R} \to \mathbb{R}$, which is known as reward shaping in the context of reinforcement learning  \cite{ng1999policy}. } Possible examples for $f$ include:
\begin{itemize}
    \item Clipping: setting $f(l_k)=\min(\beta,l_k)$ is used to deal with large loss variations and stabilize training \cite{mnih2015human}. 
    \item Baseline: setting $f(l_k)=l_k-\beta$ is called a constant baseline \cite{sutton2018reinforcement} and is  often used  to reduce  the variance of the Monte Carlo estimate of the stochastic gradient \cite{ng1999policy}. 
    \item Scaling: setting $f(l_k)=\beta l_k$ only affects the magnitude of the gradient step, but this can be compensated with methods using adaptive step sizes (including the widely used Adam optimizer \cite{Kingma2014a}). However, aggressive scaling can adversely affect the performance \cite{gu2016q,islam2017reproducibility}.
\end{itemize}

\RevA{To summarize, it has been shown that training with transformed losses, i.e., assuming $\hat{l}_k = f(l_k)$ in \eqref{eq:PolicyGradient1}, is quite robust and can even be beneficial in some cases (e.g., by reducing gradient variance through baselines). Hence, one may conclude that the training success is to a large extent determined by the relative ordering of the losses (i.e., the distinction between good actions and bad actions). In this paper, reward shaping is exploited for pre-processing before quantizing the transformed losses to a finite number of bits. }

\section{Learning with Quantized Feedback}
\label{sec:quantized}

Previous work has mostly relied on ideal feedback, where $\hat{l}_k = l_k$ \cite{Aoudia2018, Aoudia2018a, Raj2018, DeVrieze2018}. Robustness of learning with respect to additive noise according to $\hat{l}_k = l_k + n_k$, $n_k \sim \mathcal{N}(0,\sigma^2)$, was demonstrated in \cite{Goutay2018}. In this paper, we take a different view and assume that there only exists a \emph{binary feedback channel} from the receiver to the transmitter. In this case, the losses must be quantized before transmission. 

\subsection{Conventional Quantization}

\subsubsection*{Optimal Quantization} Given a distribution of the losses $p(l_k)$ and $q$ bits that can be used for quantization, the mean squared quantization error is 
\begin{align}
\label{eq:quantizer}
    \RevC{D = \mathbb{E}\{ (l_k - Q(l_k))^2\}.} 
\end{align}
With $q$ bits, there are $2^q$ possible quantization levels which can be optimized to minimize $D$, e.g., using the Lloyd-Max algorithm \cite{lloyd1982least}.

\subsubsection*{Adaptive Quantization}

\begin{figure}
\centering
\includegraphics[width=0.95\columnwidth]{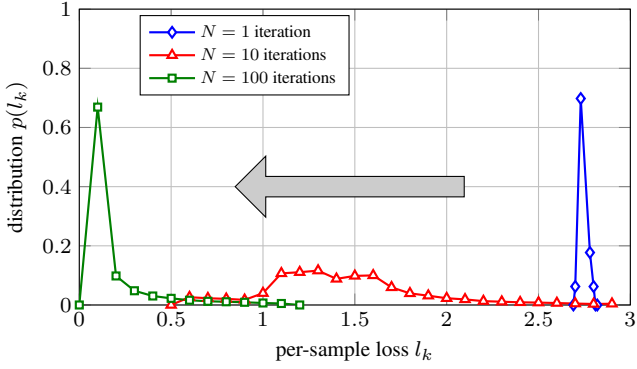}
\caption{Illustration of the non-stationary loss distribution as a function of the number of training iterations in the alternating optimization. }
\label{fig:loss_distribution}
\end{figure}

In our setting, the distribution of the per-sample losses varies over time as illustrated in Fig.~\ref{fig:loss_distribution}. For non-stationary variables, adaptive quantization can be used. The source distribution can be estimated based on a finite number of previously seen values and then adapted based on the Lloyd-Max algorithm. If the source and sink adapt based on quantized values, no additional information needs to be exchanged. If adaptation is performed based on unquantized samples, the new quantization levels need to be conveyed from the source to the sink. In either case, a sufficient number of realizations are needed to accurately estimate the loss distribution and the speed of adaptation is fixed. 

\subsubsection*{Fixed Quantization}
\label{fixed-quantization}
We aim for a strategy that does not require overhead between transmitter and receiver.  A simple non-adaptive strategy is to apply a fixed quantization. Under fixed quantization, we divide up the range  $[0,\bar{l}]$ into $2^q-1$ equal-size regions of size $\Delta = \bar{l}/2^q$ so that 
\begin{align}
    \RevC{Q(l)=  \frac{\Delta}{2} + \Delta \left\lfloor \frac{l}{\Delta} \right\rfloor.}
\end{align}Here, $\bar{l}$ is the largest loss value of interest. \RevC{The corresponding thresholds are located at $m\bar{l}/2^q$, where $m \in \{1, \ldots, 2^{q}-1\}$.} Hence, the function $Q(l)$ and its inverse $Q^{-1}(l)$ are fully determined by $\bar{l}$ and the number of bits $q$. 

\subsection{Proposed Quantization}
\label{sec:proposed_quantization}

Given the fact that losses can be transformed without much impact on the optimization, as described in Sec.~\ref{sec:losstransform}, we propose a novel strategy that employs adaptive pre-processing followed by a fixed quantization scheme. The proposed method operates on mini-batches of size $B_T$. In particular, the receiver (source) applies the following steps: 
\begin{enumerate}

    \item Clipping: we clip the losses to lie within a range $[l_{\min},l_{\max}]$. Here, $l_{\min}$ is the smallest loss in the current mini-batch, while $l_{\max}$ is chosen such that the $5\%$ largest losses in the \RevC{mini-batch} are clipped. This effectively excludes very large per-sample losses which may be regarded as outliers. We denote this operation by $f_{\text{clip}}(\cdot)$.

    \item Baseline: we then shift the losses with a fixed baseline $l_{\min}$. This ensures that all losses are within the range $[0,l_{\max}-l_{\min}]$. We denote this operation by $f_{\text{bl}}(\cdot)$. 
    \item Scaling: we scale all the losses by $1/(l_{\max}-l_{\min})$, so that they are within the range $[0,1]$. We denote this operation by $f_{\text{sc}}(\cdot)$. 
    \item Fixed quantization: 
    finally, we use a fixed quantization with $q$ bits and send $Q(\tilde{l_k})$, where \RevC{$Q(\cdot)$ is defined in \eqref{eq:quantizer} and} $\tilde{l}_k= f(l_k) = f_{\text{sc}}(f_{\text{bl}}(f_{\text{clip}}(l_k)))$, i.e., $f \define f_{\text{sc}}  \circ f_{\text{bl}} \circ f_{\text{clip}}$ denotes the entire pre-processing. \RevC{For simplicity, a natural mapping of quantized losses to bit vectors $\mathbb{B}^q$ is assumed where quantization levels are mapped in ascending order to $(0,\ldots, 0,0)^\transpose$, $(0,\ldots, 0,1)^\transpose$, \ldots, $(1,\ldots, 1,1)^\transpose$. In general, one may also try to optimize the mapping of bit vectors to the quantization levels in order to improve the robustness of the feedback transmission. }
    
\end{enumerate}

The transmitter (sink) has no knowledge of the functions $f_{\text{clip}}(\cdot)$, $f_{\text{bl}}(\cdot)$, or $f_{\text{sc}}(\cdot)$, and interprets the losses as being in the interval $[0,1]$. It thus applies $\hat{l}_k=Q^{-1}(\tilde{l}_k) \in [0,1]$ and uses the values $\hat{l}_k$ in \eqref{eq:PolicyGradient1}. \RevA{We note that some aspects of this approach are reminiscent of the Pop-Art algorithm from \cite{van2016learning}, where shifting and scaling are used to address non-stationarity during learning. In particular, Pop-Art can be used for general supervised learning, where the goal is to fit the outcome of a parameterized function (e.g., a neural network) to given targets (e.g., labels) by minimizing a loss function. Pop-Art adaptively normalizes the targets in order to deal with large magnitude variations and also address non-stationary targets. However, Pop-Art and the proposed method are different algorithms that have been proposed in different contexts, e.g., Pop-Art does not deal with quantization issues during learning. } 

\RevA{In terms of complexity overhead, the proposed method requires one sorting operation in order to identify and clip the largest losses in each mini-batch (step 1). The baseline and scaling (steps 2 and 3) can be implemented with one real addition followed by one real multiplication. Finally, the quantizer can be implemented by using a look-up table approach. At the transmitter side (sink), the method only requires the dequantization step, which again can be implemented using a look-up table. }

\subsection{Impact of Feedback Quantization}
The effect of quantization can be assessed via the Bussgang Theorem \cite{rowe1982memoryless}, which is a generalization of MMSE decomposition. If we assume $l_k \sim p(l)$ with mean $\mu_l$ and variance $\sigma^2_l$, then  \begin{align}
    Q(l_k)=g l_k + w_k,\label{eq:BussgangModel}
\end{align}
in which $g \in \mathbb{R}$ is the Bussgang gain and $w_k$ is a random variable, uncorrelated with $l_k$, provided we set 
\begin{align}
    g =\frac{\mathbb{E}\{l_k Q(l_k)\}-\mu_{l}\mathbb{E}\{Q(l_k)\}}{\sigma_{l}^{2}}. \label{eq:BussgangGain}
\end{align}
\RevC{In general, the distribution of $w_k$ may be hard (or impossible) to derive in closed form. Note that the mean of $w_k$ is $\mathbb{E}\{Q(l_k)\}- g \mu_l$ and the variance is $\mathbb{V}\{Q(l_k)\} -g^2 \sigma^2_l$. }
When the number of quantization bits $q$ increases, $Q(l_k) \to l_k$ and thus $g \to 1$. 

If we replace $l_k$ with $Q(l_k)$ in \eqref{eq:policy_gradient}, denote the corresponding gradient function by $\nabla_{\tau}\ExpL_{T}^{\mathrm{q}}(\tau)$, and substitute \eqref{eq:BussgangModel}, then  the following proposition holds. 
\begin{prop}\label{prop:1quant}
Let $\bm{\gamma}_k = l_k \nabla_{\tau}\log\pi_{\tau}(\tilde{x}_k|m_k)$, $l_k \in [0,1]$, with $\nabla_{\tau}\ExpL_{T}(\tau)=\mathbb{E}\{ \bm{\gamma}_k\}$, 
and $\bm{\gamma}^{\mathrm{q}}_k = Q(l_k) \nabla_{\tau}\log\pi_{\tau}(\tilde{x}_k|m_k)$, then 
\begin{align}
    \label{eq:Thmquant1}
    & \mathbb{E}\{ \bm{\gamma}^{\mathrm{q}}_k\} = \nabla_{\tau}\ExpL_{T}^{\mathrm{q}}(\tau) = g\nabla_{\ptx} \ExpL_T(\ptx)\\
    & \mathbb{V}\{ \bm{\gamma}^{\mathrm{q}}_k\} \le   g^{2}\mathbb{V}\{\bm{\gamma}_{k}\}+(g\bar{w}+\bar{w}^2)\mathrm{tr}\{\mathbf{J}(\tau)\}
\end{align}
where $\mathbf{J}(\tau) = \mathbb{E}\{ \nabla_{\tau}\log\pi_{\tau}(\tilde{x}_k|m_k) \nabla^\transpose_{\tau}\log\pi_{\tau}(\tilde{x}_k|m_k)\}  \succeq 0$ is the Fisher information matrix of the transmitter parameters $\tau$ and $\bar{w}=\max_{l}|gl-Q(l)| = |1-1/2^{q-1}-g|$ is a measure of the maximum quantization error.
\end{prop}
\begin{IEEEproof}
See Appendix. 
\end{IEEEproof}
\smallskip

Hence, the impact of quantization, under a sufficiently large mini-batch size is a scaling of the expected gradient. Note that this scaling will differ for each mini-batch.  The variance is affected in two ways: a scaling with $g^2$ and an additive term that depends on the maximum quantization error and the Fisher information at $\tau$. When $q$ increases, $g \to 1$ and $\bar{w} \to 0$, so that $\mathbb{V}\{ \bm{\gamma}^{\mathrm{q}}_k\}  \to \mathbb{V}\{\bm{\gamma}_{k}\}$, as expected. 

In general, the value of $g$ is hard to compute in closed form, but for 1-bit quantization and a Gaussian loss distribution, \eqref{eq:BussgangGain} admits a closed-form solution.\footnote{For Gaussian losses, $\bar{w}$ in Proposition \ref{prop:1quant} is not defined. The proposition can be modified to deal with unbounded losses.} In particular,
\begin{align}
g=
\begin{cases}
1/\sqrt{8 \pi \sigma_{l}^{2}} & \mu_l={1}/{2}\\
e^{-1/(8\sigma_{l}^{2})}/\sqrt{8 \pi \sigma_{l}^{2}} & \mu_l \in \{0,1\}.
\end{cases}
\label{eq:BussgangGain2}
\end{align}
In light of the distributions from Fig.~\ref{fig:loss_distribution}, we observe that (after loss transformation) for most iterations, $\mu_l \approx 1/2$ and $\sigma^2_l$ will be moderate (around $1/(8 \pi)$), leading to $g\approx 1$. Only after many iterations $\mu_l < 1/2$ and $\sigma^2_l$ will be small, leading to $g \ll 1$. Hence, for sufficiently large batch sizes, $1$-bit quantization should not significantly affect the learning convergence rate.

\subsection{Impact of Noisy Feedback Channels}
\label{sec:impact_of_noise}

For the proposed pre-processing and quantization scheme, distortions are introduced through the function $f(\cdot)$ (in particular the clipping) and the quantizer $Q(\cdot)$. Moreover, additional impairments may be introduced when the quantized losses are transmitted over a noisy feedback channel. We will consider the case where the feedback channel is a binary symmetric channel with flip probability $p \in [0,1/2)$. Our numerical results (see Sec.~\ref{sec:noisy_feedback_channel}) indicate that the learning process is robust against such distortions, even for very high flip probabilities. In order to explain this behavior, it is instructive to first consider the case where the transmitted per-sample losses are entirely random and completely unrelated to the training data. In that case, one finds that
\begin{align}
    &\mathbb{E}\{ \hat{l}_k \nabla_{\ptx} \log \pi_{\ptx}(\tilde{x}_k|m_k)\}= \mathbb{E}\{ \hat{l}_k \} \mathbb{E}\left\{ \nabla_{\ptx} \log \pi_{\ptx}(\tilde{x}_k|m_k)\right\} = 0 \nonumber 
\end{align}
regardless of the loss distribution or quantization scheme. 
The interpretation is that for large mini-batch sizes, random losses simply ``average out'' and the applied gradient in \eqref{eq:PolicyGradient1} is close to zero. We can exploit this behavior and make the following statement. 
\begin{prop}\label{prop:1bitnoisy}
Let $\bm{\gamma}^\mathrm{e}_k = \hat{l}_k\nabla_{\ptx} \log \pi_{\ptx}(\tilde{x}_k|m_k)$ where the binary version of $Q({l}_k)$ has been subjected to a binary symmetric channel with flip probability $p$ to yield $\hat{l}_k$. Then, for $1$-bit and $2$-bit quantization \RevC{with a natural mapping of bit vectors to quantized losses}, we have

\begin{align}
     \mathbb{E}\{ \bm{\gamma}^{\mathrm{e}}_k\} =  \nabla_{\tau}\ExpL_{T}^{\mathrm{e}}(\tau) = (1-2p)\nabla_{\ptx} \ExpL_T^{\mathrm{q}}(\ptx).
 \nonumber
\end{align}
Moreover, for $1$-bit quantization, 
\begin{align}
     \mathbb{V}\{ \bm{\gamma}^{\mathrm{e}}_k\}  \le  \mathbb{V}\{\bm{\gamma}_{k}^{\mathrm{q}}\}+4p(1-p)\Vert\nabla_{\tau}\ell_{T}^{\mathrm{q}}(\tau)\Vert^{2} +p\mathrm{tr}\{\mathbf{J}(\tau)\}.
 \nonumber
\end{align}
\end{prop}
\begin{IEEEproof}
See Appendix. 
\end{IEEEproof}
\smallskip

Hence, for a sufficiently large mini-batch size, the gradient is simply scaled by a factor $1-2p$. This means that even under very noisy feedback, learning should be possible.

\begin{rem}
Note that when using small mini-batches, the empirical gradients computed via \eqref{eq:PolicyGradient1} will deviate from the expected value $(1-2p)\nabla_{\ptx} \ExpL_T^{\text{q}}(\ptx)$: they will not be scaled exactly by $1-2p$ and they will be perturbed by the average value of $p \nabla_{\ptx} \log \pi_{\ptx}(\tilde{x}_k|m_k)$. Hence, robustness against large $p$ can only be offered for large mini-batch sizes. 
\end{rem}

\section{Numerical Results}
\label{sec:simulation}

In this section, we provide extensive numerical results to verify and illustrate the effectiveness of the proposed loss quantization scheme. In the following, the binary feedback channel is always assumed to be noiseless except for the results presented in Sec.~\ref{sec:noisy_feedback_channel}.\footnote{TensorFlow source code is available at \url{https://github.com/henkwymeersch/quantizedfeedback}.}

\subsection{Setup and Parameters}

\subsubsection{Channel Models}

We consider two memoryless channel models $p(y |x)$: the standard AWGN channel $y = x + n$, where $n\sim \mathcal{CN}(0, \sigma^2) $, and a \RevC{simplified memoryless fiber-optic channel which is defined by the recursion}

\begin{align}
    \label{eq:nlpn}
    x_{i+1} = x_{i} e^{\jmath L{\gamma}\mid x_{i}\mid ^{2}\slash K} + n_{i+1}, \quad 0\leq i < K,
\end{align}
where $x_0 = x$ is the channel input, $y = x_K$ is the channel output, $n_{i+1} \sim \mathcal{CN}(0, \sigma^2/K)$, $L$ is the total link length, $\sigma^2$ is the noise power, and ${\gamma} \geq 0$ is a nonlinearity parameter. Note that this channel reverts to the AWGN channel when ${\gamma} =0$. For our numerical analysis, we set $L = \RevC{5000}\,$km, ${\gamma} = 1.27\,$rad/W/km, $K = \RevC{50}$, and $\sigma^2 = -21.3\,$dBm, which are the same parameters as in \cite{li2018achievable, Aoudia2018a, Keykhosravi2019}. For both channels, we define $\text{SNR} \define P/\sigma^2$. Since the noise power is assumed to be fixed, the SNR is varied by varying the signal power $P$.

The model in \eqref{eq:nlpn} assumes ideal distributed amplification across the optical link and is obtained from the nonlinear
Schr\"odinger equation by neglecting dispersive effects, see, e.g., \cite{Yousefi2011a} for more details about the derivation. 
Because dispersive effects are ignored, the model does not necessarily reflect the actual channel conditions in realistic fiber-optic transmission. The main interest in this model stems from its simplicity and analytical tractability while still capturing some realistic nonlinear effects, in particular the nonlinear phase noise. The model has been studied intensively in the literature, including detection schemes \cite{Ho2005, Lau2007c, tan2011ml}, signal constellations \cite{Lau2007c, Haeger2013tcom}, capacity bounds \cite{Turitsyn2003, Yousefi2011a, keykhosravi2017tighter, Keykhosravi2019}, and most recently also in the context of machine learning \cite{li2018achievable, Aoudia2018a}. 
In the following, we refer to the model as the nonlinear phase-noise channel to highlight the fact that it should not be seen as an accurate model for fiber-optic transmission.

\subsubsection{Transmitter and Receiver Networks} Following previous work, the functions $\tx$ and $\rx$ are implemented as multi-layer neural networks. A message $m$ is first mapped to a $M$--dimensional "one-hot" vector where the $m$--th element is $1$ and all other elements are $0$. Each neuron takes inputs from the previous layer and generates an output according to a learned linear mapping followed by a fixed nonlinear activation function. The final two outputs of the transmitter network are normalized to ensure
${1}/{B}\sum_{k=1}^{B} |x_k|^2=P$, $B\in \{B_T, B_R\}$, and then used as the channel input. The real and imaginary parts of the channel observation serve as the input to the receiver network. All network parameters are summarized in Table~\ref{tab:network_parameters}, where $M = 16$.

\begin{table}
\centering
\caption{Neural network parameters, where $M = 16$}
\begin{tabular}{c|ccc|ccc}
\toprule
& \multicolumn{3}{c}{transmitter $\tx$}         & \multicolumn{3}{|c}{receiver $\rx$} \\ \midrule
layer               & 1 & 2-3  & 4      & 1     & 2-3     & 4          \\ 
number of neurons   & M & 30   & 2      & 2     & 50      & M          \\ 
activation function & - & ReLU & linear & -     & ReLU    & softmax    \\ 
\bottomrule
\end{tabular}
\label{tab:network_parameters}
\end{table}

\subsubsection{Training Procedure}

For the alternating optimization, we first fix the transmitter and train the receiver for $\iterrx = 30$ iterations with a mini-batch size of $B_R = 64$. Then, the receiver is fixed and the transmitter is trained for $\itertx = 20$ iterations with $B_T = 64$. This procedure is repeated $\iter = 4000$ times \RevC{for the AWGN channel. For the nonlinear phase-noise channel, we found that more iterations are typically required to converge, especially at high input powers, and we consequently set $\iter = 6000$.} The Adam optimizer is used to perform the gradient updates, where $\alpha_T = 0.001 $ and $ \alpha_R = 0.008$. \RevC{The reason behind the unequal number of training iterations for the transmitter and receiver is that the receiver network is slightly bigger than the transmitter network and thus requires more training iterations to converge.  }

\subsubsection{Transmitter Exploration Variance}

We found that the parameter $\sigma_p^2$ has to be carefully chosen to ensure successful training. In particular, choosing $\sigma_p^2$ too small will result in insufficient exploration and slow down the training process. On the other hand, if $\sigma_p^2$ is chosen too large, the resulting noise may in fact be larger than the actual channel noise, resulting in many falsely detected messages and unstable training. In our simulations, we use $\sigma_p^2 = P \cdot 10^{-3} $.

\begin{figure}
\centering
\includegraphics[width=0.95\columnwidth]{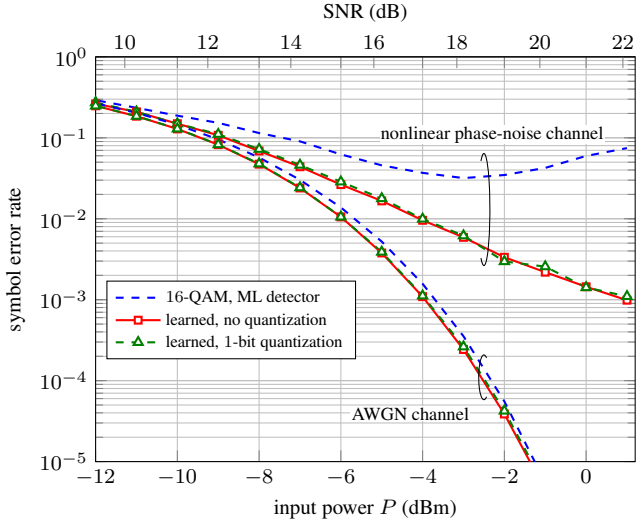}
\caption{Symbol error rate achieved for $M=16$. The training SNR is $15\,$dB for the AWGN channel, whereas training is done separately for each input power (i.e., SNR) for \RevC{the nonlinear phase-noise channel}. }
\label{fig:loss_distribution2}
\end{figure}

\begin{figure}%
\centering
\subfigure[]{%
\label{fig:first}%
\includegraphics[width=0.45\columnwidth]{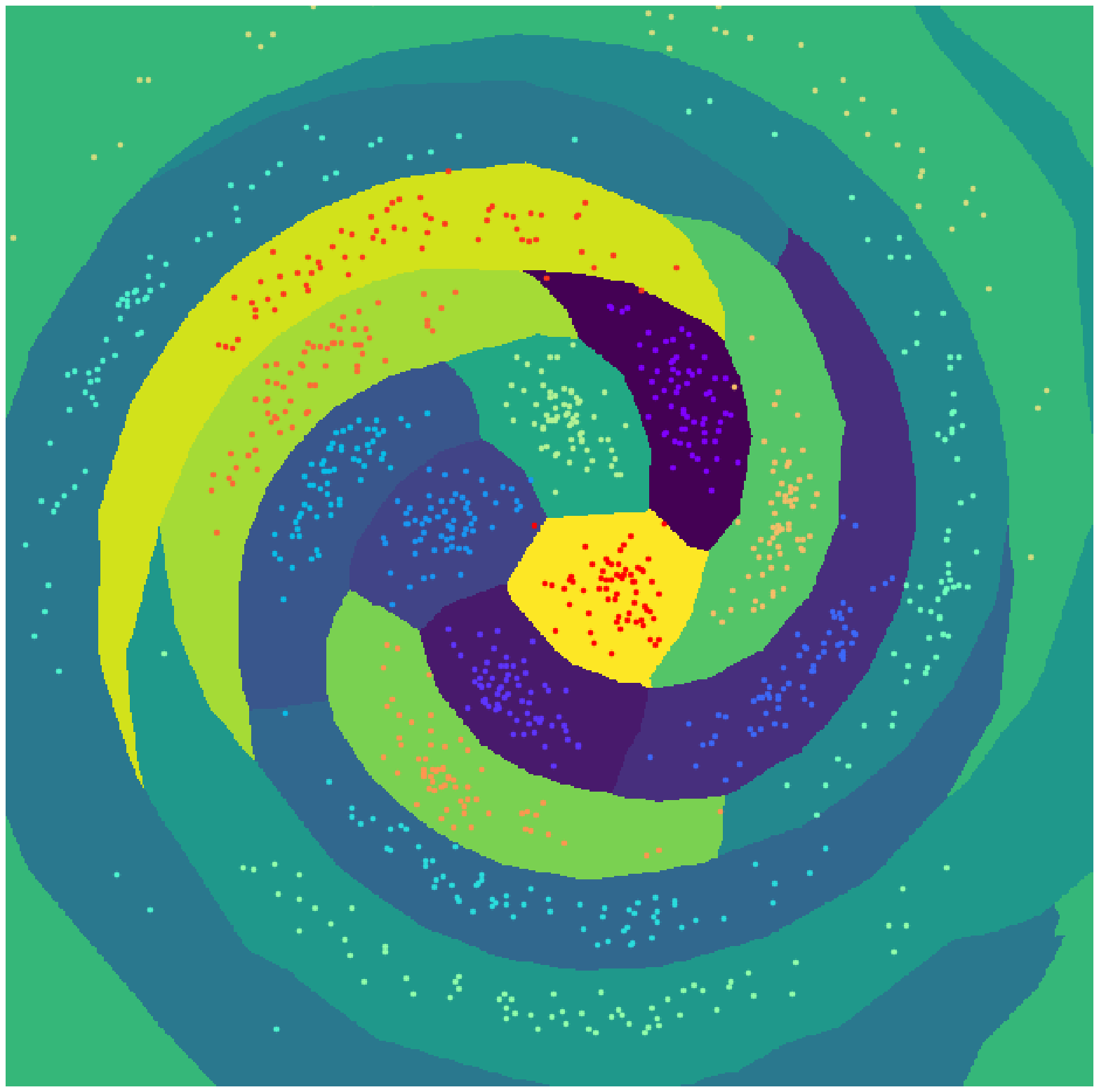}}%
\qquad
\subfigure[]{%
\label{fig:second}%
\includegraphics[width=0.45\columnwidth]{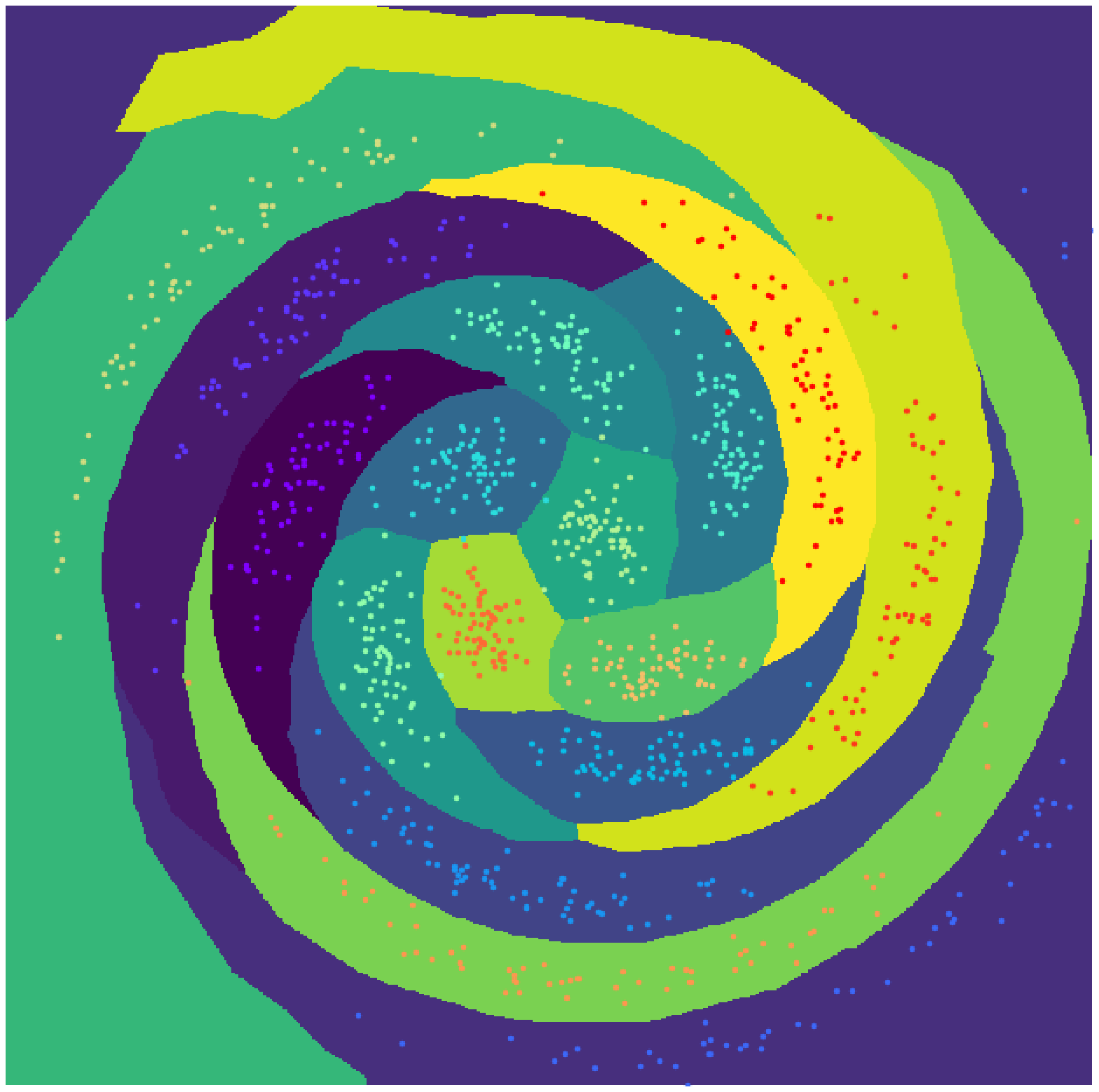}}%
\caption{Learned decision regions for \RevC{the nonlinear phase-noise channel}, $M=16$, and $P=-3\,$dBm (a) without quantizing per-sample losses and (b) using the proposed quantization scheme and 1-bit quantization.}
\label{fig: decision_region}
\end{figure}

\subsection{Results and Discussion}
\subsubsection{Perfect vs Quantized Feedback}
We start by evaluating the impact of quantized feedback on the system performance, measured in terms of the symbol error rate (SER). For the AWGN channel, the transmitter and receiver \RevC{are} trained for a fixed $\text{SNR} = 15~\text{dB}$ (i.e., $P = - 6.3~\text{dBm}$ \RevC{such that $\text{SNR} = P/\sigma^2 = - 6.3~\text{dBm} + 21.3~\text{dBm} = 15~\text{dB}$}) and then evaluated over a range of SNRs \RevC{by changing the signal power} (similar to, e.g., \cite{Aoudia2018a}). \RevC{For the nonlinear phase-noise channel}, this approach cannot be used because optimal signal constellations and receivers are highly dependent on the transmit power.\footnote{In principle, the optimal signal constellation may also depend on the SNR for the AWGN channel.} Therefore, a separate transmitter--receiver pair is trained for each input power $P$. Fig.~\ref{fig:loss_distribution2} shows the achieved SER assuming both perfect feedback without quantization and a $1$-bit feedback signal based on the proposed method. \RevC{For both channels, the resulting communication systems with $1$-bit feedback quantization have very similar performance to the scenario where perfect feedback is used for training, indicating that the feedback quantization does not significantly affect the learning process.} As a reference, the performance of standard $16$-QAM with a maximum-likelihood (ML) detector is also shown. \RevA{The ML detector makes a decision according to 
\begin{align}
    \label{eq:ml}
    \hat{x}_\text{ML} = \argmax\limits_{m \in \{1,\ldots,M\}} p(y|s_m),
\end{align}
where $s_1, \ldots, s_M$ are all constellation points. For the nonlinear phase-noise channel, the channel likelihood $p(y|x)$ can be derived in closed form, see \cite[p.~225]{Ho2005}. For the AWGN channel, \eqref{eq:ml} is equivalent to a standard minimum Euclidean-distance detector.} The learning approach outperforms this baseline for both channels, \RevB{which is explained by the fact that the transmitter neural network learns better modulation formats (i.e., signal constellations) compared to $16$-QAM.}

Fig.~\ref{fig: decision_region} visualizes the learned decision regions for the quantized (right) and unquantized (left) feedback schemes assuming \RevC{the nonlinear phase-noise channel with $P = -3\,$dBm}. Only slight differences are observed which can be largely attributed to the randomness of the training process.

\subsubsection{Impact of Number of Quantization Bits}

\begin{figure}
\centering
\includegraphics[width=0.95\columnwidth]{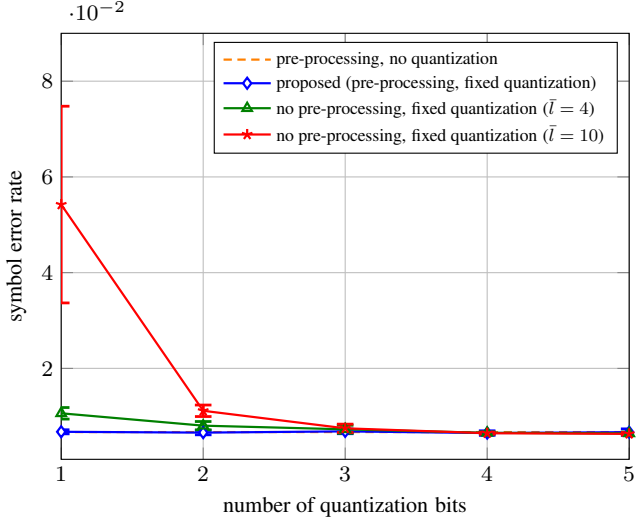}
\caption{Impact of the number of quantization bits on the achieved performance for \RevC{the nonlinear phase-noise channel} with $M =16$, \RevC{$P=-3~\text{dBm}$}. Results are averaged over $10$ different training runs where error bars indicate the standard deviation between the runs. } 
\label{fig:ser_vs_num_bits}
\end{figure}

Next, \RevC{the nonlinear phase-noise channel} for a fixed input power \RevC{$P = -3~\text{dBm}$ }is considered to numerically evaluate the impact of the number of quantization bits on the performance. Fig.~\ref{fig:ser_vs_num_bits} shows the achieved SER when different schemes are used for quantizing the per-sample losses. For a fixed quantization scheme without pre-processing (see Sec.~\ref{fixed-quantization}), the performance of the trained system is highly sensitive to the number of quantization bits and the assumed quantization range $[0, \bar{l}]$. For $\bar{l}=10$ with $1$ quantization bit, the system performance deteriorates noticeably and the training outcome becomes unstable, as indicated by the error bars (which are averaged over $10$ different training runs). For the proposed quantization scheme, the performance of the trained system is (i) essentially independent on the number of bits used for quantization and (ii) virtually indistinguishable from a system trained with unquantized feedback.

\subsubsection{Impact on Convergence Rate}

\begin{figure}
\centering
\includegraphics[width=0.95\columnwidth]{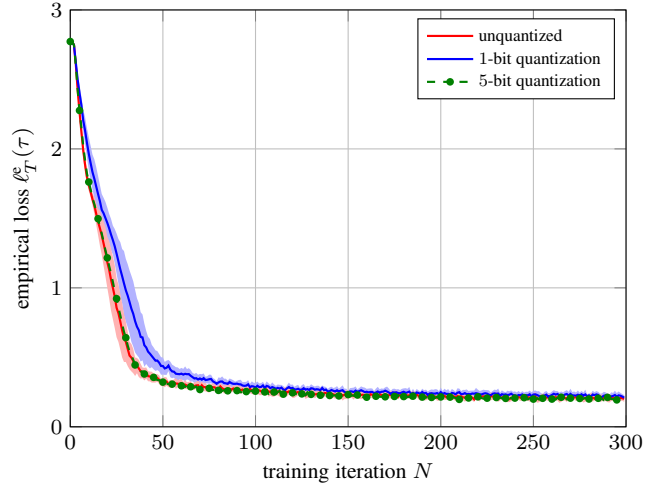}
\caption{Evolution of $\EmpL_T(\ptx)$ during the alternating optimization for \RevC{the nonlinear phase-noise channel} with $M=16$, $P=-3~\text{dBm}$. Results are averaged over $15$ different training runs where the shaded area indicates one standard deviation between the runs.}
\label{fig:loss_distribution3}
\end{figure}

In Fig.~\ref{fig:loss_distribution3}, we show the evolution of the empirical cross-entropy loss $\EmpL_T(\ptx)$ during the alternating optimization for \RevC{the nonlinear phase-noise channel with $P=-3~\text{dBm}$}. It can be seen that quantization manifests itself primarily in terms of a slightly decreased convergence rate during training. For the scenario where per-sample losses are quantized with $5$ bits, the empirical losses $\EmpL_T(\ptx)$ converged after about \RevC{$160$ iterations}, which is the same as in the case of un-quantized feedback. For $1$-bit quantization, the training converges slightly slower, after around \RevC{$200$ iterations}\RevB{, which is a minor degradation compared to the entire training time. However, the slower convergence rate implies that it is harder to deal with changes in the channel. Hence, with 1-bit quantization, the coherence time should be longer compared to with unquantized feedback.}

\begin{figure}
\centering
\includegraphics[width=0.95\columnwidth]{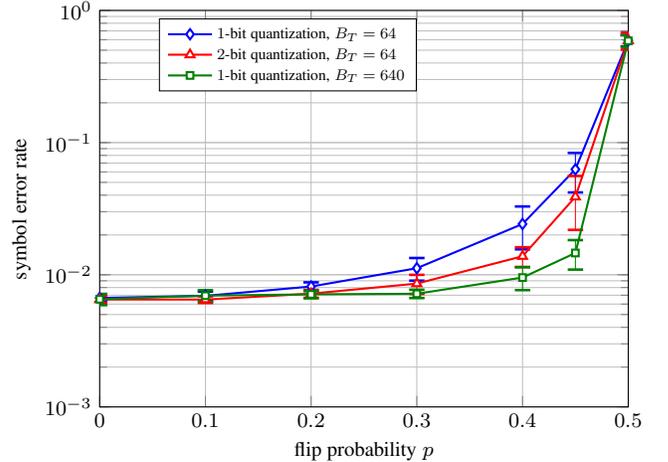}
\caption{Performance on \RevC{the nonlinear phase-noise channel} with \RevC{$M=16$, $P=-3~\text{dBm}$} when transmitting quantized losses over a noisy feedback channel modeled as a binary symmetric channel with flip probability $p$. Results are average over 10 runs where the error bars indicate one standard deviation between runs.} 
\label{fig:ser_vs_flip_probability}
\end{figure}
\subsubsection{Impact of Noisy Feedback}\label{sec:noisy_feedback_channel}

In order to numerically evaluate the effect of noise during the feedback transmission, we consider again \RevC{the nonlinear phase-noise channel} for a fixed input power\RevC{ $P = -3~\text{dBm}$}. Fig.~\ref{fig:ser_vs_flip_probability} shows the achieved SER when transmitting the quantized per-sample losses over a binary symmetric channel with flip probability $p$ (see Sec.~\ref{sec:impact_of_noise}). It can be seen that the proposed quantization scheme is highly robust to the channel noise. For the assumed mini-batch size $B_T = 64$, performance starts to decrease only for very high flip probabilities and remains essentially unchanged for $p<0.1$ with $1$-bit quantization and for $p<0.2$ with $2$-bit quantization. A theoretical justification for this behavior is provided in Proposition \ref{prop:1bitnoisy}, which states that the channel noise manifests itself only as a scaling of the expected gradient. Thus, one may also expect that the learning process can withstand even higher flip probabilities by simply increasing the mini-batch size. Indeed, Fig.~\ref{fig:ser_vs_flip_probability} shows that when increasing the mini-batch size from $B_T=64$ to $B_T=640$, the noise tolerance for $1$-bit quantization increases significantly and performance remains unchanged for flip probabilities as high as $p=0.3$. 

Note that for $p=0.5$, the achieved SER is slightly better than $(M-1)/M \approx 0.938$ corresponding to random guessing. This is because the receiver learning is still active, even though the transmitter only performs random explorations.

\section{Conclusions}
\label{sec:conclusion}

We have proposed a novel method for data-driven learning of physical-layer communication in the presence of a binary feedback channel. Our method relies on an adaptive clipping, shifting, and scaling of losses followed by a fixed quantization at the receiver, and a fixed reconstruction method at the transmitter. We have shown that the proposed method (i) can lead to good performance even under $1$-bit feedback; (ii) does not significantly affect the convergence speed of learning; and (iii) is highly robust to noise in the feedback channel. 

The proposed method can be applied beyond physical-layer communication, to reinforcement learning problems in general, and distributed multi-agent learning in particular. 






\section*{Appendix}
\subsection*{Proof of Proposition \ref{prop:1quant}}
The mean of $\bm{\gamma}^{\text{q}}_k$ can be computed as
\begin{align*}
& \mathbb{E}\{ \bm{\gamma}^{\text{q}}_k\} = \nabla_{\tau}\ExpL_{T}^{\text{q}}(\tau) \nonumber \\
& =\mathbb{E}\{Q(l_k)\nabla_{\tau}\log\pi_{\tau}(\tilde{x}_k|m_k)\}\\
 & =g\mathbb{E}\{l_k\nabla_{\tau}\log\pi_{\tau}(\tilde{x}_k|m_k)\}+\mathbb{E}\{w_k\nabla_{\tau}\log\pi_{\tau}(\tilde{x}_k|m_k)\}\\
 & =g\mathbb{E}\{l_k\nabla_{\tau}\log\pi_{\tau}(\tilde{x}_k|m_k)\}+\mathbb{E}\{w_k\}\mathbb{E}\{\nabla_{\tau}\log\pi_{\tau}(\tilde{x}_k|m_k)\}\\
 & =g\mathbb{E}\{l_k\nabla_{\tau}\log\pi_{\tau}(\tilde{x}_k|m_k)\}=g\nabla_{\tau}\ell_{T}(\tau).
\end{align*}
We have made use of the fact that $w_k$ is uncorrelated with $l_k$ and that \eqref{eq:exp_grad_log_policy} holds. The variance can similarly be bounded as follows:
\begin{align*}
    & \mathbb{V}\{ \bm{\gamma}^{\text{q}}_k\} \nonumber \\
    & =
    \mathbb{E}\{(Q(l_{k}))^{2}\Vert\nabla_{\tau}\log\pi_{\tau}(\tilde{x}_{k}|m_{k})\Vert^{2}\}-g^{2}\Vert\nabla_{\tau}\ell_{T}(\tau)\Vert^{2}
    \\
    & =g^{2}\mathbb{E}\{l_{k}^{2}\Vert\nabla\log\pi_{\tau}(x_{k}|m_{k})\Vert^{2}\}-g^{2}\Vert\nabla_{\tau}\ell_{T}(\tau)\Vert^{2}\\
    & +\mathbb{E}\{w_{k}^{2}\Vert\nabla\log\pi_{\tau}(x_{k}|m_{k})\Vert^{2}\}\\
    & +2\mathbb{E}\{gl_{k}w_{k}\Vert\nabla\log\pi_{\tau}(x_{k}|m_{k})\Vert^{2}\}\\
    & \le g^{2}\mathbb{V}\{\bm{\gamma}_{k}\}+\bar{w}^2\text{tr}\{\mathbf{J}(\tau)\}\\
    & -2g\mathbb{E}\{w_{k}l_{k}\Vert\nabla\log\pi_{\tau}(\tilde{x}_{k}|m_{k})\Vert^{2}\}\\ \nonumber
    & \le g^{2}\mathbb{V}\{\bm{\gamma}_{k}\}+\bar{w}^2\text{tr}\{\mathbf{J}(\tau)\}  +2g\bar{w}\text{tr}\{\mathbf{J}(\tau)\} \nonumber
\end{align*}
We have made use of $-w_{k} l_{k} =l_k (g l_k - Q(l_k)) \le \max_{l_k} |g l_k-Q(l_k)| = \bar{w}$, that $l_k \le 1$, and that $\text{tr}\{\mathbf{J}(\tau)\} = \mathbb{E}\{\Vert\nabla\log\pi_{\tau}(x_{k}|m_{k})\Vert^{2}\}$. 
\subsection*{Proof of Proposition \ref{prop:1bitnoisy}}

For the proposed adaptive pre-processing and fixed $1$-bit quantization, the quantized losses $l_k$ are either $\Delta/2=1/4$ or $1-\Delta/2=3/4$. Assuming transmission over the binary symmetric channel, the gradient in \eqref{eq:policy_gradient} can be written as 
\begin{align*}
    \nabla_{\ptx} \ExpL_T^{\text{e}}(\ptx) = \mathbb{E}\{ Q(l_{k})^{1-n_k} (1-Q(l_{k}))^{n_k} \nabla_{\ptx} \log \pi_{\ptx}(\tilde{x}_k|m_k)\},
\end{align*}
where $n_k$ are independent and identically distributed Bernoulli random variables with parameter $p$. Since $n_k$ is independent of all other random variables, we can compute
\begin{align*}
\mathbb{E}[Q(l_{k})^{1-n_k} (1-Q(l_{k}))^{n_k} \,|\, Q(l_{k})] = (1-2p) Q(l_{k}) + p.
\end{align*} 
Hence, 
\begin{align*}
    \label{eq:noisy_gradient3}
    & \mathbb{E}\{ \bm{\gamma}^{\text{e}}_k\}= \nabla_{\ptx} \ExpL_T^{\text{e}}(\ptx)\\
    & = \mathbb{E}\{ ((1-2p) Q(l_{k}) + p) \nabla_{\ptx} \log \pi_{\ptx}(\tilde{x}_k|m_k)\}\\
    & = (1-2p) \mathbb{E}\{ Q(l_{k}) \nabla_{\ptx} \log \pi_{\ptx}(\tilde{x}_k|m_k)\}+ p  \mathbb{E}\{ \nabla_{\ptx} \log \pi_{\ptx}(\tilde{x}_k|m_k)\} \nonumber \\
    & = (1-2p)\nabla_{\ptx} \ExpL_T^{\text{q}}(\ptx),
\end{align*}
where the last step follows from \eqref{eq:exp_grad_log_policy}. For 2-bit quantization, the possible values are $\Delta/2=1/8$ (corresponding to bits 00), $3\Delta/2=3/8$ (corresponding to 01), $1-3\Delta/2=5/8$ (corresponding to 10),  $1-\Delta/2=7/8$ (corresponding to 11). It then follows that when the transmitted loss is $Q(l_{k})$, the received loss is
\begin{align*}
 Q(l_{k})    & \text{ with prob. }(1-p)^2\\
 1-Q(l_{k}) & \text{ with prob. }p^2\\
 \text{other}& \text{ with prob. }p(1-p)
\end{align*}
so that the expected received loss is $(1-2p)Q(l_{k})+p$. 

The variance under 1-bit quantization can be computed as
\begin{align*}
    & \mathbb{V}\{\bm{\gamma}_{k}^{\text{e}}\}\\
    &=\mathbb{E}\{(\bm{\gamma}_{k}^{\text{e}})^{2}\}-(1-2p)^{2}\Vert\nabla_{\tau}\ell_{T}^{\text{q}}(\tau)\Vert^{2}\\
    & = \mathbb{E}\{(Q(l_{k}))^{2(1-n_{k})}(1-Q(l_{k}))^{2n_{k}}\Vert\nabla_{\tau}\log\pi_{\tau}(\tilde{x}_{k}|m_{k})\Vert^{2}\}\\
    & -(1-2p)^{2}\Vert\nabla_{\tau}\ell_{T}^{\text{q}}(\tau)\Vert^{2}\\
    & = \mathbb{E}\{Q^{2}(l_{k})\Vert\nabla_{\tau}\log\pi_{\tau}(\tilde{x}_{k}|m_{k})\Vert^{2}\}+p\mathbb{E}\{\Vert\nabla\log\pi_{\tau}(\tilde{x}_{k}|m_{k})\Vert^{2}\}\\
    & -2p\mathbb{E}\{Q(l_{k})\Vert\nabla\log\pi_{\tau}(\tilde{x}_{k}|m_{k})\Vert^{2}\}-(1-2p)^{2}\Vert\nabla_{\tau}\ell_{T}^{\text{q}}(\tau)\Vert^{2}\\
    & = \mathbb{V}\{\bm{\gamma}_{k}^{\text{q}}\}+4p(1-p)\Vert\nabla_{\tau}\ell_{T}^{\text{q}}(\tau)\Vert^{2}+p\text{tr}\{\mathbf{J}(\tau)\}\\
     & -2p\mathbb{E}\{Q(l_{k})\Vert\nabla_{\tau}\log\pi_{\tau}(\tilde{x}_{k}|m_{k})\Vert^{2}\}\\
     & \le \mathbb{V}\{\bm{\gamma}_{k}^{\text{q}}\}+4p(1-p)\Vert\nabla_{\tau}\ell_{T}^{\text{q}}(\tau)\Vert^{2}+p \text{tr}\{\mathbf{J}(\tau)\}, 
\end{align*}
where the last step holds since $Q(l_k) \ge  0$.

\bibliographystyle{IEEEtran}
\bibliography{references} 

\begin{thebibliography}{10}
\providecommand{\url}[1]{#1}
\csname url@samestyle\endcsname
\providecommand{\newblock}{\relax}
\providecommand{\bibinfo}[2]{#2}
\providecommand{\BIBentrySTDinterwordspacing}{\spaceskip=0pt\relax}
\providecommand{\BIBentryALTinterwordstretchfactor}{4}
\providecommand{\BIBentryALTinterwordspacing}{\spaceskip=\fontdimen2\font plus
\BIBentryALTinterwordstretchfactor\fontdimen3\font minus
  \fontdimen4\font\relax}
\providecommand{\BIBforeignlanguage}[2]{{%
\expandafter\ifx\csname l@#1\endcsname\relax
\typeout{** WARNING: IEEEtran.bst: No hyphenation pattern has been}%
\typeout{** loaded for the language `#1'. Using the pattern for}%
\typeout{** the default language instead.}%
\else
\language=\csname l@#1\endcsname
\fi
#2}}
\providecommand{\BIBdecl}{\relax}
\BIBdecl

\bibitem{Samuel2017}
N.~Samuel, T.~Diskin, and A.~Wiesel, ``{Deep MIMO Detection},'' in \emph{Proc.
  IEEE Int. Workshop on Signal Processing Advances in Wireless Communications
  (SPAWC)}, 2017.

\bibitem{Nachmani2018}
E.~Nachmani, E.~Marciano, L.~Lugosch, W.~J. Gross, D.~Burshtein, and Y.~Be'ery,
  ``{Deep Learning Methods for Improved Decoding of Linear Codes},'' \emph{IEEE
  J. Sel. Topics Signal Proc.}, vol.~12, no.~1, pp. 119--131, Feb. 2018.

\bibitem{OShea2017}
T.~O'Shea and J.~Hoydis, ``{An Introduction to Deep Learning for the Physical
  Layer},'' \emph{IEEE Trans. on Cognitive Communications and Networking},
  vol.~3, no.~4, pp. 563--575, Dec. 2017.

\bibitem{Doerner2018}
S.~D{\"{o}}rner, S.~Cammerer, J.~Hoydis, and S.~ten Brink, ``{Deep
  Learning-Based Communication Over the Air},'' \emph{IEEE J. Sel. Topics
  Signal Proc.}, vol.~12, no.~1, pp. 132--143, Feb. 2017.

\bibitem{karanov2018end}
B.~Karanov, M.~Chagnon, F.~Thouin, T.~A. Eriksson, H.~B{\"u}low, D.~Lavery,
  P.~Bayvel, and L.~Schmalen, ``End-to-end deep learning of optical fiber
  communications,'' \emph{J. Lightw. Technol.}, vol.~36, no.~20, pp.
  4843--4855, 2018.

\bibitem{li2018achievable}
S.~Li, C.~H{\"a}ger, N.~Garcia, and H.~Wymeersch, ``Achievable information
  rates for nonlinear fiber communication via end-to-end autoencoder
  learning,'' in \emph{Proc. European Conf. Optical Communication (ECOC)},
  Rome, Italy, 2018.

\bibitem{Jones2018}
R.~T. Jones, T.~A. Eriksson, M.~P. Yankov, and D.~Zibar, ``{Deep Learning of
  Geometric Constellation Shaping including Fiber Nonlinearities},'' in
  \emph{Proc. European Conf. Optical Communication (ECOC)}, Rome, Italy, 2018.

\bibitem{Lee2018}
H.~Lee, I.~Lee, and S.~H. Lee, ``{Deep learning based transceiver design for
  multi-colored VLC systems},'' \emph{Opt. Express}, vol.~26, no.~5, pp.
  6222--6238, Mar. 2018.

\bibitem{OShea2018}
T.~J. O'Shea, T.~Roy, and N.~West, ``{Approximating the Void: Learning
  Stochastic Channel Models from Observation with Variational Generative
  Adversarial Networks},'' \emph{arXiv:1805.06350}, 2018.

\bibitem{Ye2018}
H.~Ye, G.~Y. Li, B.-H.~F. Juang, and K.~Sivanesan, ``{Channel Agnostic
  End-to-End Learning based Communication Systems with Conditional GAN},''
  \emph{arXiv:1807.00447}, 2018.

\bibitem{Aoudia2018}
F.~A. Aoudia and J.~Hoydis, ``{End-to-End Learning of Communications Systems
  Without a Channel Model},'' \emph{arXiv:1804.02276}, 2018.

\bibitem{Aoudia2018a}
------, ``{Model-free Training of End-to-end Communication Systems},''
  \emph{arXiv:1812.05929}, 2018.

\bibitem{DeVrieze2018}
C.~de~Vrieze, S.~Barratt, D.~Tsai, and A.~Sahai, ``{Cooperative Multi-Agent
  Reinforcement Learning for Low-Level Wireless Communication},''
  \emph{arXiv:1801.04541}, 2018.

\bibitem{Raj2018}
V.~Raj and S.~Kalyani, ``{Backpropagating Through the Air: Deep Learning at
  Physical Layer Without Channel Models},'' \emph{IEEE Commun. Lett.}, vol.~22,
  no.~11, pp. 2278--2281, Nov. 2018.

\bibitem{Goutay2018}
M.~Goutay, F.~A. Aoudia, and J.~Hoydis, ``{Deep Reinforcement Learning
  Autoencoder with Noisy Feedback},'' \emph{arXiv:1810.05419}, 2018.

\bibitem{Kim2018b}
M.~Kim, W.~Lee, J.~Yoon, and O.~Jo, ``{Building Encoder and Decoder with Deep
  Neural Networks: On the Way to Reality},'' \emph{arXiv:1808.02401}, 2018.

\bibitem{Tang2018}
Z.-L. Tang, S.-M. Li, and L.-J. Yu, ``{Implementation of Deep Learning-based
  Automatic Modulation Classifier on {FPGA} {SDR} Platform},''
  \emph{Electronics}, vol.~7, no.~7, p. 122, 2018.

\bibitem{Teng2018}
C.-F. Teng, C.-H. Wu, K.-S. Ho, and A.-Y. Wu, ``{Low-complexity Recurrent
  Neural Network-based Polar Decoder with Weight Quantization Mechanism},''
  \emph{arXiv:1810.12154}, 2018.

\bibitem{Fougstedt2018ecoc}
C.~Fougstedt, C.~H{\"{a}}ger, L.~Svensson, H.~D. Pfister, and
  P.~Larsson-Edefors, ``{{ASIC} Implementation of Time-Domain Digital
  Backpropagation with Deep-Learned Chromatic Dispersion Filters},'' in
  \emph{Proc. European Conf. Optical Communication (ECOC)}, Rome, Italy, 2018.

\bibitem{Aoudia2019}
F.~A. Aoudia and J.~Hoydis, ``{Towards Hardware Implementation of Neural
  Network-based Communication Algorithms},'' \emph{arXiv:1902.06939}, 2019.

\bibitem{ng1999policy}
A.~Y. Ng, D.~Harada, and S.~J. Russell, ``Policy invariance under reward
  transformations: Theory and application to reward shaping,'' in
  \emph{Proceedings of the Sixteenth International Conference on Machine
  Learning}.\hskip 1em plus 0.5em minus 0.4em\relax Morgan Kaufmann Publishers
  Inc., 1999, pp. 278--287.

\bibitem{mnih2015human}
V.~Mnih, K.~Kavukcuoglu, D.~Silver, A.~A. Rusu, J.~Veness, M.~G. Bellemare,
  A.~Graves, M.~Riedmiller, A.~K. Fidjeland, G.~Ostrovski \emph{et~al.},
  ``Human-level control through deep reinforcement learning,'' \emph{Nature},
  vol. 518, no. 7540, p. 529, 2015.

\bibitem{sutton2018reinforcement}
R.~S. Sutton and A.~G. Barto, \emph{Reinforcement learning: An
  introduction}.\hskip 1em plus 0.5em minus 0.4em\relax MIT press, 2018.

\bibitem{Kingma2014a}
D.~P. Kingma and J.~Ba, ``{Adam: A Method for Stochastic Optimization},'' in
  \emph{Proc. ICLR}, 2015.

\bibitem{gu2016q}
S.~Gu, T.~Lillicrap, Z.~Ghahramani, R.~E. Turner, and S.~Levine, ``Q-prop:
  Sample-efficient policy gradient with an off-policy critic,'' \emph{arXiv
  preprint arXiv:1611.02247}, 2016.

\bibitem{islam2017reproducibility}
R.~Islam, P.~Henderson, M.~Gomrokchi, and D.~Precup, ``Reproducibility of
  benchmarked deep reinforcement learning tasks for continuous control,''
  \emph{arXiv preprint arXiv:1708.04133}, 2017.

\bibitem{lloyd1982least}
S.~Lloyd, ``Least squares quantization in pcm,'' \emph{IEEE transactions on
  information theory}, vol.~28, no.~2, pp. 129--137, 1982.

\bibitem{van2016learning}
H.~P. van Hasselt, A.~Guez, M.~Hessel, V.~Mnih, and D.~Silver, ``Learning
  values across many orders of magnitude,'' in \emph{Advances in Neural
  Information Processing Systems}, 2016, pp. 4287--4295.

\bibitem{rowe1982memoryless}
H.~Rowe, ``Memoryless nonlinearities with {Gaussian} inputs: Elementary
  results,'' \emph{The BELL system technical Journal}, vol.~61, no.~7, pp.
  1519--1525, 1982.

\bibitem{Keykhosravi2019}
K.~Keykhosravi, G.~Durisi, and E.~Agrell, ``{Accuracy Assessment of
  Nondispersive Optical Perturbative Models through Capacity Analysis},''
  \emph{Entropy}, vol.~21, no.~8, pp. 1--19, aug 2019.

\bibitem{Yousefi2011a}
M.~I. Yousefi and F.~R. Kschischang, ``{On the per-sample capacity of
  nondispersive optical fibers},'' \emph{IEEE Trans. Inf. Theory}, vol.~57,
  no.~11, pp. 7522--7541, November 2011.

\bibitem{Ho2005}
K.-P. Ho, \emph{{Phase-modulated Optical Communication Systems}}.\hskip 1em
  plus 0.5em minus 0.4em\relax Springer, 2005.

\bibitem{Lau2007c}
A.~P. Lau and J.~M. Kahn, ``{16-QAM Signal Design and Detection in Presence of
  Nonlinear Phase Noise}.''\hskip 1em plus 0.5em minus 0.4em\relax IEEE, July
  2007, pp. 53--54.

\bibitem{tan2011ml}
A.~S. Tan, H.~Wymeersch, P.~Johannisson, E.~Agrell, P.~Andrekson, and
  M.~Karlsson, ``An ml-based detector for optical communication in the presence
  of nonlinear phase noise,'' in \emph{2011 IEEE International Conference on
  Communications (ICC)}.\hskip 1em plus 0.5em minus 0.4em\relax IEEE, 2011, pp.
  1--5.

\bibitem{Haeger2013tcom}
C.~H{\"{a}}ger, A.~{Graell i Amat}, A.~Alvarado, and E.~Agrell, ``{Design of
  {APSK} Constellations for Coherent Optical Channels with Nonlinear Phase
  Noise},'' \emph{IEEE Trans. Commun.}, vol.~61, no.~8, pp. 3362--3373, August
  2013.

\bibitem{Turitsyn2003}
K.~S. Turitsyn, S.~A. Derevyanko, I.~V. Yurkevich, and S.~K. Turitsyn,
  ``{Information Capacity of Optical Fiber Channels with Zero Average
  Dispersion},'' vol.~91, no.~20, p. 203901, nov 2003.

\bibitem{keykhosravi2017tighter}
K.~Keykhosravi, G.~Durisi, and E.~Agrell, ``A tighter upper bound on the
  capacity of the nondispersive optical fiber channel,'' in \emph{2017 European
  Conference on Optical Communication (ECOC)}.\hskip 1em plus 0.5em minus
  0.4em\relax IEEE, 2017, pp. 1--3.

\end{thebibliography}
\label{references}

\end{document}